\documentstyle[11pt,emulateapj,apjfonts,epsf]{article}
 
\def\kms{\relax \ifmmode {\,\rm km\,s}^{-1}\else \,km\,s$^{-1}$\fi}

\def\farcs{\hbox{$.\!\!^{\prime\prime}$}}

\def\arcsec{\hbox{$^{\prime\prime}$}}
\def\secd#1.#2{ #1\farcs#2 }               
\def\H2{$\rm H_2$}
\def\Bg{$\rm Br\gamma$}
\def\Hb{$\rm H\beta$}
\def\ergcms{{\rm erg}\,{\rm cm}^{-2}\,{\rm s}^{-1}} 
 
\received{}
\accepted{}
 
\lefthead{{\H2} and {\Bg} emission in bipolar PNe}
\righthead{Guerrero et al.}

\begin{document}

\title{{\H2} and {\Bg} narrow-band imaging of bipolar planetary 
       nebulae.}

\author{M. A. Guerrero\altaffilmark{1}, 
                                                                               E.
                                                                               Villaver}
                                                                               
\affil{Instituto de Astrof\'{\i}sica de Canarias, E--38200
       La Laguna (Tenerife), Spain}

\author{A. Manchado}
\affil{Instituto de Astrof\'{\i}sica de Canarias, E--38200
       La Laguna (Tenerife), Spain}
\affil{Consejo Superior de Investigaciones Cient\'{\i}ficas, CSIC}

\author{P. Garc\'{\i}a-Lario}
\affil{ISO Data Centre, Estaci\'on de Villafranca del Castillo, 
       Apdo. Postal 50727, 
       E-28080 Madrid, Spain}

\author{F. Prada\altaffilmark{2}}
\affil{Instituto de Astrof\'{\i}sica de Canarias, E--38200
       La Laguna (Tenerife), Spain}

\altaffiltext{1}{Currently at Department of Astronomy, University of 
       Illinois at Urbana-Champaign, 1002 West Green Street, Urbana, 
       IL 61801}
\altaffiltext{2}{Currently at Centro Astron\'omico Hispano-Aleman,
                 Apdo 511, 
                 E-04080, Almer\'{\i}a, Spain}

\submitted{{\it To appear in} {\rm The Astrophysical Journal Suplement Series}}

\begin{abstract}

We present near-IR narrow-band continuum-subtracted images in the 
{\H2} $2.122\mu$m, and {\Bg} $2.166\mu$m emission lines for a sample 
of 15 bipolar planetary nebulae. 
{\H2} emission was definitely detected for most of the objects in this sample 
(13 out of 15). 
The very high {\H2} detection rate supports the idea that bipolar planetary 
nebulae have important reservoirs of molecular material and offer suitable 
physical conditions for the excitation of {\H2}. 
The strength of the {\H2} emission and the {\H2}/{\Bg} flux ratio are found 
to correlate with the morphology of the bipolar nebulae observed. 
Bipolar PNe with broad and bright rings exhibit stronger {\H2} emission 
than bipolar PNe with narrow twists.
High-quality (sub-arcsec) [N~{\sc ii}] and H$\alpha$ optical images have 
been used to compare the distribution of the ionized and molecular 
material. 
The {\H2} emission lies just outside the optical [N~{\sc ii}] emission zone.

\end{abstract}


\section{Introduction}

Planetary Nebulae (PNe) form when low- and intermediate-mass ($<10$ 
M$_\odot$) stars evolve off the asymptotic giant branch (AGB) phase 
to white dwarfs. 
In this short-lived transition, AGB stars eject most of their envelope 
which become thick circumstellar shells. 
The central star will later ionize this material and a new PN arises. 
While the visible PN is dominated by ionized gas, molecular material may 
still survive in high-density clumps or in the photo-dissociation region 
(PDR) where the optical depth to ionizing photons is high (Tielens 1993). 
The presence of residual molecular gas in PNe, remnant of the AGB envelope, 
provides us with important hints about the mass-loss history and geometry on 
and at the end of the AGB, and offers some insight into the formation 
mechanism of PNe.

Extensive surveys in the CO (2--1) line in PNe (Huggins \& Healy 1989; 
Huggins et al.\ 1996, and references therein) have shown that molecular 
envelopes are widely observable in PNe and constitute a significant 
component of their masses. 
The large format near-IR arrays available nowadays make the observations 
in the {\H2} 2.122~$\mu$m emission line (hereinafter we will refer to this 
line as {\H2}, except when explicitly indicated) a powerful tool for the 
study of the distribution of molecular gas in PNe. 
Since the first detection of this line in the young PN NGC\,7027 (Treffers 
et al.\ 1976), over 60 other PNe and proto-PNe have been detected in this 
line (Webster et al.\ 1988; Zuckerman \& Gatley 1988; Kastner et al.\ 1994, 
1996, and references therein; Hora \& Latter 1996), and the number of 
detections is increasing steadily. 
Some doubts about the validity of the {\H2} emission line as a quantitative 
probe of the total molecular mass have been posed ({\it e. g.} Dinerstein 
1991) as this line can be excited either by fluorescence through the 
absorption of UV photons from the central star in the photo-dissociation 
front (Black \& van Dishoeck 1987; Sternberg \& Dalgarno 1989) or by shocks 
({\it e. g.} Burton, Hollenbach, \& Tielens 1992). 
The comparison between CO and {\H2} emission in PNe (Huggins \& Healy 1989; 
Huggins et al.\ 1996) shows that the {\H2} emission is indeed related to 
the amount of molecular gas in PNe.

The CO and {\H2} observations of PNe have revealed that molecular gas is 
predominantly detected in bipolar PNe, although emission is also found 
among other morphological classes. 
Bipolar PNe have relatively dense equatorial regions which provide a safe 
haven for molecules due to self- and dust-shielding of UV photons. 
They are also associated to high-mass progenitor stars (Peimbert \& 
Torres-Peimbert 1983; Zuckerman \& Gatley 1988; Corradi \& Schwarz 1995; 
Huggins et al.\ 1996) which presumably have experienced higher mass 
ejections.

The bipolar PN class is not homogeneous as a whole, as two distinct 
types of detailed morphologies exist (Manchado et al.\ 1996, hereafter 
M96; Manchado 1997): 
(1) bipolar PNe with a compact core or narrow waist (class B); 
(2) bipolar PNe with a broad ring structure in the waist (class Br). 
Balick (1987) already noticed this morphological difference (the so-called 
early and late butterfly types) and proposed that B PNe evolve into Br PNe.

Webster et al.\ (1988) reported observations of a sample of 11 bipolar PNe 
and noted that Br PNe have the highest {\H2} intensity and {\H2} to {\Bg} 
flux ratio. 
However, their results should be used with caution due to poor spatial 
resolution ($\sim 5\arcsec\times14\arcsec$) and sampling. 
The latter tends to maximize the {\H2} to {\Bg} flux ratio. 
This can be seen in some individual objects, as NGC\,7027, whose flux 
ratio, as given by Webster et al., does not account for the {\H2} to 
{\Bg} large ratio variations over the nebula (Cox et al.\ 1997).

In order to gain more insigth into the correlation between {\H2} emission 
and the detailed morphology of bipolar PNe, this paper presents near-IR 
narrow-band images in the {\H2} and {\Bg} lines of a sample of 15 bipolar 
B and Br PNe selected from the narrow-band images displayed by M96. 
The {\H2} and {\Bg} fluxes of these PNe have been measured and accurate 
{\H2} intensities and {\H2} to {\Bg} flux ratios of each subgroup of 
bipolar PNe have been obtained. 
The results are compared with these previously presented by Webster et 
al.\ (1988) and the implications on the formation and evolution of 
bipolar PNe are discussed. 
Moreover, the comparison of the {\H2} and optical images from the M96 catalog 
have been used to analyze the relative distribution of ionized and molecular
material.
Finally, a detailed description of the individual morphology of some peculiar 
PNe in the sample is given.

\section{Observations}

\subsection{Sample selection criteria}

The nebulae were selected according to the following criteria: 
\begin{enumerate}
\item{All the nebulae are bipolar or quadrupolar PNe, showing a 'waist' 
      and (at least) two bipolar lobes. }
\item{
      They have dereddened H$\beta$ fluxes in the range $3.2\,10^{-13} \le 
      {\rm F(H\beta)} \le 5.1\,10^{-11}\;\ergcms$, and [N~{\sc ii}] 
      morphologies indicative of the presence of dense low-ionization 
      equatorial regions. } 
\item{They have angular sizes larger than $15\arcsec$ to resolve them 
      properly, and smaller than $150\arcsec$ to make them fit into the 
      field of view. }
\end{enumerate}

\subsection{Data acquisition}

Near-IR images of the selected fields were obtained on 24 and 25 
June 1996 with the MPI f\"ur Astronomie General-Purpose Infrared Camera 
(MAGIC) attached to the f/8 Cassegrain focus of the Calar Alto (CAHA) 2.2m 
telescope. 
The detector was a Rockwell $256\times256$ pixel$^2$ NICMOS3 array. 
The high resolution imaging mode was used with a pixel scale of 
$0.64\arcsec$ pixel$^{-1}$, and a field of view of $164\arcsec$. 
The filter parameters are described in Table~1.

Different observing strategies were followed for compact and extended sources. 
Compact sources were observed rastering them to different locations of the 
detector, so that the image resulting from the combination of different 
frames before centering can be used as a sky frame.
In the case of extended sources, adjacent blank sky regions observed 
between consecutive on-source exposures were combined to obtain the 
corresponding sky frame. 
In both cases, the integration time on the object was always 100 sec. 
The images presented here are composed of many shorter exposures which 
were then individually sky-subtracted and flat-fielded before centering and 
averaging them to produce the final images. 
The FWHM of point sources in the final images, which includes both the seeing 
at $\lambda = 2.2\mu$m and instrumental effects, are in the range 
$1\farcs1-1\farcs5$. 
The weather was photometric through the whole observing run. 
The sensitivity at $3\sigma$ over the background was found to be 
$\sim~3.0\,10^{-5}\;\ergcms\,{\rm sr}^{-1}$.
%
%

\subsection{Data reduction}

As opposed to optical emission from PNe, strongly dominated by narrow 
nebular emission lines, the situation is very different in the near-IR,
where the continuum emission sometimes becomes very important and can even 
dominate the observed emission.
Unfortunately, there were no \H2 and {\Bg} continuum filters available at the 
telescope. 
Thus, in order to subtract the continuum emission from the narrow-band 
images, we obtained additional broad-band $K^\prime$ images.

The subtraction of the continuum emission requires some detailed 
explanation. 
First, we scaled the broad- and narrow-band emission using images of the 
flux standard star SA\,107--1006 (Casali \& Hawarden 1992) obtained at both
filters, 
assuming that the continuum of this star is roughly constant through the 
$K^\prime$-band. 
Then, for each nebula we subtracted the emission from the {\H2} and {\Bg} 
to the continuum image of the nebula. 
Finally, after scaling, we subtracted these non-contaminated by emission 
lines continuum images to the original \H2 and {\Bg} images in order to 
obtain continuum-subtracted images in these lines. 
The {\H2} and {\Bg} narrow-band continuum-subtracted images for all the PNe 
are presented in Figure~1. 
For comparison, the [N~{\sc ii}] optical images from M96 are also presented 
at the same scale.

The contribution from other emission lines to the flux observed in the 
$K^\prime$-band has carefully been considered. 
Besides {\Bg} and {\H2} 2.122 $\mu$m, the most prominent emission lines 
in this band are Br8 H~{\sc i}, He~{\sc i} 2.058, and a few {\H2} emission 
lines at 1.957, 2.033, 2.223 and 2.248 $\mu$m. 
After re-scaling the continuum image, the expected contribution from 
Br8 H~{\sc i} and He~{\sc i} 2.058 is estimated to be $\sim2\%$ the 
intensity of {\Bg}, whereas the contribution of the {\H2} lines is 
$\leq5\%$ the intensity of the {\H2} 2.122 $\mu$m line.

Similarly, we have analyzed the contribution of the He~{\sc i} $2.1126\mu$m 
line to the flux observed in our narrow-band {\H2} images. 
Hora, Latter, \& Deutsch (1999) presented reliable measurements of the intensity
of 
the He~{\sc i} $2.1126\mu$m and {\Bg} lines for 33 PNe. 
The average intensity of the He~{\sc i} $2.1126\mu$m line is $\sim5\%$ 
the intensity of the {\Bg} line. 
Therefore, a significant contribution from the He~{\sc i} $2.1126\mu$m line 
to our narrow-band {\H2} images is only expected when {\Bg} is much more 
intense than {\H2}. 
Even in these cases, the emission in the {\H2} and He~{\sc i} lines comes 
from different regions in the PN, and therefore it is still possible to 
distinguish between {\H2} and He~{\sc i} emission.

The standard star SA\,107--1006 was used for flux calibration. 
The measured {\H2} and {\Bg} fluxes are given in Table~2. 
Given the inherent high-noise and short-term fluctuations of the near-IR 
sky background, and the uncertainties above described affecting the continuum 
subtraction, the measured fluxes can be affected by relative errors which 
have been estimated to be $\le5\%$ for fluxes larger than 
$1.0\,10^{-12}\;\ergcms$, and $5-20 \%$ for smaller fluxes.

\section{Detection of molecular hydrogen}

Table~2 lists the usual names\footnote{
The acronym `He' is used through this work to name Henize's objects, 
despite the fact that `Hen' is the acronym used in astronomical
dictionaries. 
The former is used in The Strasbourg-ESO Catalogue of Galactic Planetary
Nebulae (Acker et al.\ 1992) and is widely accepted.}
and approximate galactic coordinates (the 
PN G name as described by Acker et al.\ 1992) of the objects observed (columns 
1 and 2); 
the morphological classification (column 3); 
the angular size of the whole nebula and its ring (columns 4, and 5) in the 
optical as given by M96; 
and the dereddened {\Hb} fluxes (column 6) as obtained from Acker et al. 
(1992). 
Table~2 also summarizes the main results of this work: the measured fluxes in 
the \H2 and {\Bg} lines (columns 7 and 8); the \H2 to {\Bg} flux ratio (column 
9); and the \H2 surface brightness ratio between the lobes and the ring (column 
10), $f_{\rm LR}$, defined as
\begin{equation}
f_{\rm LR} = \frac{B_{\rm lobe}}{B_{\rm ring}}
\end{equation}
where $B_{\rm lobe}$ and $B_{\rm ring}$ is a rough estimate of the peak 
surface brightness of the lobes and ring, respectively.

Emission in the {\H2} images (Fig.~1) was detected above $3\sigma$ 
level in all but one (He 2-428) of the nebulae in our sample. 
The possible contamination due to the He~{\sc i} 2.1126 $\mu$m line may 
become relevant only for the objects with the lowest {\H2} to {\Bg} ratios: 
M~1-57, M~1-59, M~2-46, and He~2-437. 
In particular, we cannot exclude that the emission detected in M~2-46 
is due to He~{\sc i}. 
The detection of {\H2} emission is confirmed in M~1-57, M~1-59, and He~2-437 
according to morphological criteria; 
the {\H2} emission is detected in the lobes of those nebulae, where {\Bg} 
emission is not detected. 
In the case of He~2-437, where {\Bg} emission is detected in the lobes, it
should be noted that the position angle of the emission in the \H2 image 
differs from that in the {\Bg} image, but it follows the [N~{\sc ii}] 
emission. 
Since the {\H2} fluxes of M~1-59 and M~2-46 given in table~2 correspond to the 
whole nebulae, these values should be considered as upper limits; the {\H2} 
fluxes of M~1-57 and He~2-437 were measured in the lobes, excluding the 
central part. 
Therefore, the emission detected in 13 of the bipolar PNe in our sample is 
connected with {\H2} emission from these nebulae.

Figure~2-[{\it top}] shows the dereddened {\Bg} fluxes plotted 
against the {\Hb} fluxes. 
Adopting case B recombination, $T_{\rm e}\,=\,10^4\,K$ and $N_{\rm e}\,=\,
10^3\,{\rm cm}^{-3}$, the theoretical {\Bg} to {\Hb} flux ratio is expected 
to be $\sim~0.03$ (Osterbrock 1989). 
This relation is shown as a dashed straight line in Fig.~2-[{\it
top}]. 
Although a correlation between {\Bg} and {\Hb} fluxes is found, most of the 
objects in our sample fall below the predicted {\Bg} to {\Hb} fluxes ratio; 
this is probably due to the poor accuracy of the {\Hb} flux determination, 
based for most of the objects on a rough extrapolation from {\Hb} flux 
determination on a small portion of the object (Acker et al. 1991) which is 
likely underestimating the real {\Hb} flux.

The measured {\H2} fluxes are plotted against the {\Bg} fluxes in 
Fig.~2-[{\it bottom}]. 
It is clear that our sample breaks down into two different locations on 
this diagram. 
In agreement with the values of the {\H2} to {\Bg} flux ratios listed in 
Table~2, we distinguish two well defined subgroups in our sample: 
those PNe whose {\H2} flux is several times ($\ge 5$) stronger than the 
{\Bg} flux, and those in which the {\Bg} emission clearly dominates. 
Globally considered, the $F({\rm H_2})/F({\rm Br}\gamma)$ ratios observed 
span two orders of magnitude.
As we discuss below, these nebulae have different morphological and 
physical properties.

\subsection{{\H2} dominated bipolar PNe}

We can include BV~1, K~3-34, K~3-58, K~4-55, M~1-75, M~2-52, 
M~4-14, and M~4-17 in this group. 
In all these cases the emission observed in the \H2 line is more than 5 
times greater than the {\Bg} emission. 
The measured \H2 flux is in the range 
$5\,10^{-13}$ to $4\,10^{-12} {\ergcms}$, with peak intensities 
$1.2-5.3\times10^{-4}{\ergcms}\,{\rm sr}^{-1}$.

The morphologies observed in the {\H2} line (Fig.~1) closely 
resemble
those shown in the optical [N~{\sc ii}] lines (M96). 
The {\H2} emission is mainly concentrated in a bright ring, the ``waist'' 
of the bipolar nebula.
The peak emission of the ring has a surface brightness more than two times 
brighter than the projected edge of the lobes, as listed in Table~2. 
The surface brightness contrast between the ring and lobes is, however, much 
less extreme than in the optical [N~{\sc ii}] $\lambda 6583$ line. 
In contrast, the much fainter {\Bg} emission is detected only in the central 
ring, which appears sharper in the {\H2} line than in the {\Bg} line.

All of these PNe share some common properties. 
First, all of them exhibit well defined rings in the light of [N~{\sc ii}]. 
Indeed, they all are Br or ``quadrupolar'' (Q) with ring PNe according to 
the scheme derived from optical observations proposed by M96. 
Note that BV~1 can also be included in this group, although its ring is not 
so evident, as it is seen edge-on (Kaler, Chu, \& Jacoby 1988; Josselin et al.\ 
1999). 
The rings of these PNe are characterized by a high surface brightness 
contrast between the ring itself and its innermost part. 
The emission in the [N~{\sc ii}] and {\H2} lines arises mainly in both cases 
from the ring, but the center of the ring does not show any emission in these 
lines or very little. 
These PNe have also relatively low density (100 to 1000 cm$^{-3}$) (Acker et 
al.\ 1992), and the averaged size of their rings is 0.22 pc. 
In addition, their central stars are all faint to be detected in the optical 
images with the exception of the very faint central star of M~4-14.

\subsection{{\Bg} dominated bipolar PNe}

In contrast with the PNe above cited, the bipolar PNe He~2-437, M~1-57, 
M~1-59, NGC~6881, and PC~20 are characterized by a small {\H2} to {\Bg} 
flux ratio (F({\H2})/F({\Bg})$\sim0.1-0.5$). 
All of them show emission in the {\Bg} line strongly concentrated in the 
central, bright core. 
{\H2} emission is detected both at the central regions and along the bipolar 
lobes in most cases, although the {\H2} emission is always fainter than the 
{\Bg} emission.

He~2-437 and M~1-57 show an unresolved core. 
They also exhibit the most extremely low {\H2} to {\Bg} flux ratios 
($\sim 0.1-0.2$), whereas for M~1-59, NGC~6881, and PC~20, the {\H2} 
to {\Bg} flux ratios increase up to $0.2-0.5$. 
These latter nebulae, classified as Br PNe by M96, have small sized central 
rings, with a small surface brightness contrast between the innermost part 
of the ring and its edge in the [N~{\sc ii}] line. 
M~1-59, NGC~6881, and PC~20, therefore, show intermediate properties between 
the {\Bg} emission dominated class B PNe with unresolved cores (He~2-437 and
M~1-57) and the {\H2} emission dominated class Br PNe described above.

The bipolar PNe in this subgroup exhibit different properties than the 
{\H2} dominated bipolar PNe. 
Their central regions have a high surface brightness, in agreement with 
their larger densities ($10^3$ to $10^4$ cm$^{-3}$) (Acker et al.\ 1992). 
Their averaged size is only 0.076 pc. 
Finally, many of the central stars of this subgroup of PNe are very bright 
in the optical images.

\subsection{Comments on individual objects}

{\bf NGC 6881.} 

The [N~{\sc ii}] image of this nebula (Guerrero \& Manchado 1998) shows a 
highly collimated bipolar structure with a size of {$29\farcs5 \times 
5\arcsec$}. 
Each lobe ends as a bright knot where the emission in the [N~{\sc ii}] 
light is enhanced. 
A second, brighter pair of inner lobes extends over {$16\arcsec$}. 
Finally, a loop-like structure is located $11\arcsec$ towards the southeast 
extreme of the nebula.

The differences between the [N~{\sc ii}] and \H2 images of this nebula are 
obvious. 
Unlike observed in the optical, a pair of very opened lobes is being 
collimated by a marked $4\arcsec$ waist which roughly corresponds to 
the ring observed in the optical. 
The lobes extend $20\arcsec$ to the north-west and $13\arcsec$ to the 
south-east, almost coinciding with the south-east loop-like structure. 
As the lobes separate from the central waist, their width increases to a 
maximum dimension of {$16\arcsec$}. 
As it is illustrated in Figure~3, the corresponding width 
of the optical lobes at this position is less than {$5\arcsec$}.
Beyond the limit of the northwest lobe, but at the same position angle,  
extended emission is found as a continuation of this lobe. 
The maximum angular distance at which we found emission is {$\sim50\arcsec$}, 
almost twice the angular dimension of the ionized emission. 
NGC\,6881 is also a member of the group of peculiar PNe whose {\H2} 
morphology does not trace the ionized emission, as it is the case of J\,900 
(Shupe et al.\ 1995) and NGC\,2440 (Latter et al.\ 1995; Latter \& Hora 1997).

{\bf M 1-57}

{\H2} emission is marginally detected $\sim30\arcsec$ from the centre of 
the nebula with the same orientation of the $15\arcsec$ sized optical 
lobes (M96). 
Deeper images, both in the [N~{\sc ii}] and {\H2} lines, are needed to 
confirm this result.

{\bf He~2-428.}

This is the only bipolar PNe not detected in the \H2 line. 
It is an intriguing situation, since it belongs to the Br class of M96 and 
it is similar in many aspects to other nebulae in this group, whose {\H2} 
flux is a few times stronger than the {\Bg} emission. 
However, while {\Bg} emission is detected in He\,2-428, the nebula is not 
detected in {\H2}. 
In addition, this is the only case of a nebula in our sample in which a 
bright central star is found. 
Since the central stars of nebulae with bright {\H2} emission in our sample 
are very faint and difficult to detect, it is urged to investigate in detail 
the nature of He\,2-428 and its central star.

\section{Molecular and ionized material distribution}

In order to make a fair comparison of the distribution of the molecular 
and ionized material in these PNe, their [N~{\sc ii}] and H$\alpha$+[N~{\sc 
ii}] optical images from M96 were rebinned to the same pixel scale 
of the near-IR images ($0\farcs64$~pixel$^{-1}$). 
A gaussian filter was used to degrade the optical images and make the 
FWHM of point sources in the field similar for both sets of images 
($1\farcs1-1\farcs5$). 
The contour plot of {\H2} emission has been overlaid over a gray-scale 
representation of the [N~{\sc ii}] images in Fig.~1-[{\it bottom}].

Most of the {\H2} morphological features closely trace the corresponding 
[N~{\sc ii}] optical features. 
The bright condensations observed in the rings in {\H2} 
are also found in the [N~{\sc ii}] images. 
Furthermore, the optical design of the lobes in [N~{\sc ii}] is also 
outlined in {\H2}. 
The {\H2} emission in the bipolar lobes is relatively enhanced compared to 
the optical [N~{\sc ii}] lines. 
The $f_{\rm LR}$ ratios given in Table~2 are in the range $0.2-0.4$; 
the corresponding ratios in the [N~{\sc ii}] lines are $\sim5$ times 
smaller.

We extracted spatial profiles from the {\H2} and [N~{\sc ii}] images 
to quantify the dimensions of the molecular and low-ionization material 
distributions. 
The spatial line profiles, extracted along representative directions of the 
nebulae, are shown in Figure~3. 
They have been used to measure the size of well defined features in both 
the {\H2} and [N~{\sc ii}] images. 
In particular, the size of the rings measured from peak to peak are listed 
in Table~3. 
Given the pixel size and the expected FWHM for a point sources, the position 
is accurate within $0\farcs1-0\farcs2$. 
The {\H2} emission from the ring closely embraces the [N~{\sc ii}] optical 
emission in the {\H2} dominated bipolar PNe. 
Differences in size goes from $0\farcs6$ to $3\farcs9$. 
Furthermore, emission in the {\H2} extends further than in the [N~{\sc ii}] 
line. 
When lobes are mapped, they appear to be more prominent in the {\H2} line 
({\it e. g.} K~4-55, or M~1-75), partly due to the enhancement of {\H2} 
emission from the lobes described above. 
A singular case is the {\H2} emission detected in the lobes of NGC~6881 
described in $\S3.3$. 
In this case, we measured the lobes width at two different slit positions.

Now, it would be desirable to transform the angular differences in size 
to linear sizes. 
However, the use of statistical distances is uncertain in these objects, 
since its particular morphology makes the spherical symmetry assumption 
on which those methods are based unreasonable (Pottasch 1984; Maciel 1997). 
As there are no individual distance estimates for most of these objects, 
we have adopted the averaged value among those included in Acker et al. 
(1992), and the more recent estimations given by Van de Steene \& Ziljstra 
(1994), and Zhang (1995). 
Using these averaged distances, as given in Table~3, the thickness of the 
region located between the [N~{\sc ii}] emission peak and the {\H2} peak 
(column 7) result to be in a narrow range from $7\times10^{15}$ to 
$6.4\times10^{16}$ cm. 
Thus, the {\H2} emission occurs in a narrow region just outside the ionized 
zone.

\section{Discussion}

We have searched for molecular hydrogen emission in a sample of 15
bipolar PNe.
The high detection rate ($\sim90\%$) found in our sample is the result 
of its bias to bipolar morphologies, since the frequency of detections 
in a morphologically unbiased sample of PNe (e. g. Kastner et al.\ 1996) 
is rather small, $\sim40\%$. 
Moreover, almost $70\%$ of the combined sample of bipolar PNe surveyed by 
Kastner et al.\ (1996) and this work exhibit {\H2} emission. 
These numbers support the idea that bipolar PNe are, at least partially, 
ionization-bounded, with important reservoirs of molecular material, and 
that they offer suitable physical conditions for the excitation of {\H2}.

The {\H2} emission closely traces the [N~{\sc ii}] ionized emission in 
most of the PNe in our sample. 
This behavior has been reported, although based on mostly `qualitative' 
arguments (Kastner et al.\ 1996), data of poor ($5\arcsec - 10\arcsec$) 
spatial resolution (Beckwith, Persson, \& Gatley 1978; Webster et al.\ 
1988), or the quantitative analysis of a few large PNe, as it is the case 
of the elliptical PN NGC\,6720 (Zuckerman \& Gatley 1988).
The improved spatial resolution ($\sim1\arcsec$) of our narrow-band 
[N~{\sc ii}] and {\H2} images has enabled us to make a detailed analysis 
of the relative distribution of the ionized and {\H2} molecular emission 
in eight bipolar PNe. 
The {\H2} emission is found in a thin shell ($\le 0.1$ times the optical 
size), just outside the ionized [N~{\sc ii}] zone both in the rings and 
the walls of the lobes. 
As pointed out by Kastner et al.\ (1996), the {\H2} emission is brightest 
toward the ring of the nebula. 
Moreover, our data reveal that the {\H2} emission is enhanced in the lobes 
in comparison to the [N~{\sc ii}] optical emission.

One of the main results derived from this work is that a firm connection 
can now be established between the detailed nebular morphology of bipolar 
PNe and the {\H2} intensities and {\H2}/{\Bg} flux ratios observed. 
The PNe with Br morphology show larger {\H2} to {\Bg} flux ratios, as well 
as larger {\H2} intensities. 
In contrast, B PNe have the smallest {\H2} to {\Bg} flux ratios and {\H2} 
intensities. 
This result strengthens the earlier suggestion by Webster et al. (1988) 
that strong {\H2} emission is found almost exclusively among Br PNe. 
Adding Webster et al.'s sample to ours, we find that $\sim80\%$ of the 
Br PNe surveyed belong to the group with strong {\H2} intensities.

Why is {\H2} intensity enhanced in Br PNe? 
In order to assess and clarify the underlying physics of this correlation, 
several hypotheses have been considered, although all of them have their 
own share of difficulties. 
Moreover, they do not exclude each other.

{\it ``B and Br PNe are at different evolutionary stages''}.-- 
This idea, originally proposed by Balick (1987), is supported by the 
different physical properties of B and Br PNe described in $\S3.1$ 
and $\S3.2$. 
The Br PNe smaller densities and larger sizes suggest larger dynamical 
ages. 
Indeed, the high surface brightness of the ring of Br PNe requires a 
long time in order to allow the fast wind from the central star to 
have swept up over the ring the material that once was inside it. 
To this effect, the fact that the central stars of Br PNe are faint is 
consistent with an evolved stage.

The {\H2} to {\Bg} ratio is expected to increase with time for PNe 
from massive progenitors (Natta \& Hollenbach 1998). 
According to this model, the predicted ratios increase from $\sim0.1$ 
for young PNe to $\sim10$ for old PNe, in reasonable agreement with the 
observed ratios. 
However, Bobrowsky \& Zipoy (1989) and Natta \& Hollenbach (1998) 
models predict a steep decrease of the {\H2} intensity with time. 
For massive progenitors (0.696 M$_{\odot} < M_{\rm core} < 0.836 
{\rm M}_{\odot}$) with a high density molecular shell ($\ge 10^5$ 
cm$^{-3}$), Natta \& Hollenbach's model predicts the decline of the 
{\H2} intensity by a factor $>100$ in the early nebular evolution. 
The agreement between the theoretical predictions and the increase of 
{\H2} intensity with time is possible if a lower initial density of the 
molecular envelope ($\le 10^4$ cm$^{-3}$) is assumed. 
In this case, it has been shown (Vicini et al.\ 1999) that the intensity 
of the {\H2} 2.122 $\mu$m line does not decline for a PN with core mass 
0.7 M$_{\odot}$ as it evolves, but it is enhanced (the reader is referred 
to this paper for further details). 
If the detailed bipolar morphology is indeed related to the evolution 
of bipolar PNe, the bulk of observational data presented by Webster et 
al.\ (1988) and this paper indicate that low initial densities of the 
molecular envelopes are the rule rather than the exception in bipolar 
PNe.

{\it ``The {\H2} emission in B and Br PNe have different prevalent 
excitation mechanisms''}.-- 
Note that this does not necessarily exclude the previous evolutionary 
hypothesis, as the prevalent excitation mechanism may be different as 
the PN evolves.

Shocks may play a significant role in the excitation of molecular hydrogen 
in bipolar PNe. 
According to Natta \& Hollenbach (1998), the shock contribution to the 
intensity of the {\H2} 2.122 $\mu$m line is expected to be 
\begin{equation}
\sim0.1\times \dot{M}_{\rm RG}f_{\rm RG}/10^{-5}{\rm M}_{\odot}{\rm yr}^{-1}
\end{equation} 
over most of the PN lifetime, where $\dot{M}_{\rm RG}$ and $f_{\rm RG}$ 
are the mass-loss rate and filling factor of the red giant wind. 
Therefore, for high mass-loss rates and/or non-isotropic red giant winds, 
shock excitation would dominate the {\H2} excitation. 
Indeed, the {\H2} 2.122 $\mu$m peak intensities predicted by shock models 
(Draine, Roberge, \& Dalgarno 1983; Burton et al.\ 1992) for the typical 
expansion velocities ($30-100\kms$) of bipolar PNe (Corradi \& Schwarz 
1995) are in agreement with the observed peak intensities in our sample 
($1.2-5.3 \times 10^{-4}\,{\ergcms}\,{\rm sr}^{-1}$).

However, no evidence for shock excitation is found in the archetypical Br 
PN NGC\,2346. 
The prevalent {\H2} excitation mechanism in this PN has been shown to be 
UV excitation (Hora et al.\ 1999; Vicini et al.\ 1999). 
It can be argued that even with this evidence, the shock excitation cannot 
easily be disregarded, as the presence of fast winds and dissociating 
shocks are able to produce UV photons in the shocked wind which would 
indirectly excite the {\H2}.

On the other hand, even if shock excitation is important in bipolar PNe, 
it is not clear why {\H2} intensity should be enhanced in Br PNe and not 
in B PNe. 
Both subsamples of nebulae exhibit the same kinematical properties: the 
expansion velocity of the lobes of these bipolar PNe ranges from 30 to 
100 {\kms}, whereas rings expand more slowly, at $5-25$ {\kms} (Guerrero 
\& Manchado, unpublished data). 
Alternatively, the {\H2} line intensities may strongly depend on the density 
of the material and on the shock structure (Shull \& Hollenbach 1978; Burton 
et al.\ 1992), so that the characteristic geometry of Br PNe would 
favor the enhancement of the {\H2} emission.

An additional excitation mechanism that must be taken into account in the 
late evolution of the PN is the presence of X-rays from the hot central 
star. 
Natta \& Hollenbach (1998), and Vicini et al.\ (1999) have highlighted the 
importance of the X-ray heating in enhancing the predicted {\H2} intensity, 
specially for PNe with low initial densities of the molecular envelope. 
However, the uncertainty in the far UV and soft X-ray spectrum of PNe 
forces to use this result with special caveat. 
No X-ray emission has been detected in bipolar PNe up to now, most likely 
because of the high extinction and low luminosity of the central stars of 
bipolar PNe (Guerrero, Chu, \& Gruendl 1999).

Obviously, a detailed spatially-resolved spectroscopic analysis of the
{\H2} emission line spectrum in a large sample of bipolar PNe is needed 
to fully address these issues.

{\it ``B and Br PNe proceed from different evolutionary tracks''}.-- 
If Br PNe follow a different evolutionary track with enhanced mass-loss 
rates on the AGB, they would have much more massive molecular envelopes 
than B PNe. 
Subsequently, the {\H2} emission in Br PNe would be stronger than in 
B PNe. 
The evolution from more massive stars than the progenitors of B PNe or 
from a binary system undergoing a common envelope phase may be the cause 
for the enhanced mass-loss rate in Br PNe. 
The lack of observational evidence to support these particular evolutionary 
tracks for B and Br PNe would make it a highly speculative conclusion.

\section{Conclusions}

We have presented continuum-subtracted flux-calibrated {\H2} and 
{\Bg} images of a sample of 15 bipolar PNe. 
The high {\H2} detection rate within this sample strengthens the idea that 
bipolar PNe are molecular rich objects where {\H2} finds suitable physical 
conditions for its excitation.

The {\H2} intensity and {\H2} to {\Bg} flux ratio are found to change with 
the detailed morphology of the bipolar PNe; PNe with Br morphology exhibit 
the highest {\H2} to {\Bg} flux ratios, and {\H2} absolute fluxes. 
Although we cannot discard the possibility of B and Br PNe proceeding 
from different evolutionary tracks, most observational data indicate that 
the observed correlation reflects the fact that B and Br PNe are in a 
different evolutionary stage. 
In order to explain the enhancement of the {\H2} intensity with time, the 
initial density of the molecular envelope must be low. 
The different {\H2} intensities and {\H2} to {\Bg} flux ratios observed 
in each of these two subgroups of bipolar PNe may also be connected to 
different prevalent excitation mechanisms, which can, at the same time, 
be a consequence of the different evolutionary stage.

The {\H2} emission from {\H2}-dominated bipolar PNe is mainly located in the 
ring, although fainter {\H2} emission is also found in the lobes. 
The comparison between {\H2} and [N~{\sc ii}] images of these PNe reveals 
that the {\H2} emission is distributed in a narrow region just outside the 
ionized material and that it is more enhanced in the lobes compared to the 
ionized [N~{\sc ii}] emission.

\acknowledgments

The 2.2m telescope at the German-Spanish Astronomical Centre, Calar Alto, is 
operated by the Max-Planck-Institut f\"ur Astronomie, Heidelberg, jointly with 
the Spanish National Comission for Astronomy. 
This research was partially funded through grant PB94-1274 from the 
Direcci\'on General de Investigaci\'on Cient\'{\i}fica y T\'ecnica 
of the Spanish Ministerio de Educaci\'on y Ciencia.
MAG is supported partially by the Direcci\'on General de Ense\~nanza 
Superior e Investigaci\'on Cient\'{\i}fica of the Spanish Ministerio de 
Educaci\'on y Cultura.

\clearpage

\onecolumn

\begin{deluxetable}{lcc}
\tablenum{1}
\tablewidth{20pc}
\tablecaption{Filter description}
\tablehead{
\multicolumn{1}{c}{Filter name}& 
\multicolumn{1}{c}{Central wavelength} & 
\multicolumn{1}{c}{FWHM} \\
\multicolumn{1}{c}{} & 
\multicolumn{1}{c}{[$\mu$m]} & 
\multicolumn{1}{c}{[$\mu$m]} }
 
\startdata
H$_2$ S(1) 1--0 & 2.122 & 0.021 \nl
Br$\gamma$      & 2.166 & 0.022 \nl
$K^{\prime}$    & 2.100 & 0.340 \nl
\enddata
\end{deluxetable}

\begin{deluxetable}{llcccrrrrr}
\tablenum{2}
\tablewidth{42pc}
\tablecaption{Properties of the bipolar planetary nebula sample}
\tablehead{
\multicolumn{1}{c}{Object}        & 
\multicolumn{1}{c}{PN G}           & 
\multicolumn{1}{c}{Morph. class$^1$} & 
\multicolumn{1}{c}{PN size$^1$}      & 
\multicolumn{1}{c}{Ring size$^1$}    &
\multicolumn{1}{c}{F(H$\beta$)$^2$}  & 
\multicolumn{1}{c}{F(H$_2$)}      &
\multicolumn{1}{c}{F(Br$\gamma$)} &
\multicolumn{1}{c}{$\frac{{\rm F(H}_2)}{{\rm F(Br}\gamma)}$} &
\multicolumn{1}{c}{$f_{\rm LR}$}  \\
\multicolumn{1}{c}{} &
\multicolumn{1}{c}{} & 
\multicolumn{1}{c}{} &
\multicolumn{1}{c}{[arcsec]} &
\multicolumn{1}{c}{[arcsec]} & 
\multicolumn{3}{c}{[$10^{-14}\;{\rm erg}\,{\rm cm}^{-2}\,{\rm s}^{-1}$]} &
\multicolumn{1}{c}{} &
\multicolumn{1}{c}{}}

\startdata
M 1-57	 & $022.1-02.4$ & B &24.3&10.6&   4100 &     21 &    239 &     0.1& 0.14   \nl
M 1-59	 & $023.9-02.3$ & Br&24.2&8.9 &   5100 &     61 &    320 &     0.2& 0.22   \nl
M 2-46	 & $024.8-02.7$ & Q &30.0&7.0 &    830 &     11 &     64 &     0.2&$\dots$ \nl
PC 20	 & $031.7+01.7$ & Br&11.8&6.5 &$\dots$ &     53 &     90 &     0.6& 0.75   \nl
M 4-14	 & $043.0-03.0$ & Q & 27 &8.5 &    720 &    192 &     18 &    10.8& 0.30   \nl
He 2-428 & $049.4+02.4$ & Br& 63 & 18 &     56 &$\dots$ &     43 &$\dots$ &$\dots$ \nl
K 3-34	 & $059.0+04.6$ & Br&20.8&10.5&     32 &     47 &$\dots$ &$\dots$ & 0.37   \nl
He 2-437 & $061.3+03.6$ & B & 45 &4.6 &    780 &     12 &    115 &     0.1&$\dots$ \nl
M 1-75	 & $068.8-00.0$ & Br&11.8&6.5 &    230 &    389 &     68 &     5.7& 0.38   \nl
K 3-58	 & $069.6-03.9$ & Br& 23 &12.7&    150 &    227 &     7.2&    32.0& 0.20   \nl
NGC 6881 & $074.5+02.1$ & Q &29.5&5.0 &   2700 &    125 &    243 &     0.5& 0.19   \nl
M 4-17	 & $079.6+05.8$ & Br& 24 & 20 &    230 &    324 &     30 &    10.7& 0.28   \nl
K 4-55	 & $084.2+01.1$ & Br& 71 &9.0 &    120 &    245 &     17 &    14.0& 0.29   \nl
M 2-52	 & $103.7+00.4$ & Br& 60 & 23 &   3800 &    198 &     23 &     8.6& 0.18   \nl
BV 1     & $119.3+00.3$ & Br& 48 & 7  &    280 &    197 &$\dots$ &$\dots$ &$\dots$ \nl
\enddata			         
\tablerefs{(1) Manchado et al 1996; (2) Acker et al 1992} 
\end{deluxetable}		         

\begin{deluxetable}{lcccccc}
\tablenum{3}
\tablewidth{38pc}
\tablecaption{Comparison of the H$_2$ and [N ii] spatial distributions}
\tablehead{
\multicolumn{1}{l}{Object}         & 
\multicolumn{1}{c}{Position angle} &
\multicolumn{1}{c}{H$_2$ size}    &
\multicolumn{1}{c}{[N~{\sc ii}] size}    &
\multicolumn{1}{c}{Relative size H$_2$/[N~{\sc ii}]}  & 
\multicolumn{1}{c}{Distance} & 
\multicolumn{1}{c}{Thickness} \\
\multicolumn{1}{c}{} &
\multicolumn{1}{c}{[deg]} & 
\multicolumn{1}{c}{[arcsec]} &
\multicolumn{1}{c}{[arcsec]} & 
\multicolumn{1}{c}{} & 
\multicolumn{1}{c}{[kpc]} &
\multicolumn{1}{c}{[$10^{16}$ cm]}}

\startdata
M 1-59	 &        	    ~~40 &  ~3.2 &  ~3.2                 & 1.00 & 3.0 &  $<0.3$ \nl
M 4-14	 &        	     140 &  ~6.4 &  ~5.8                 & 1.11 & 5.2 & ~~~2.3  \nl
M 1-75	 &        	    ~~54 &  ~9.6 &  ~8.3                 & 1.15 & 3.4 & ~~~3.3  \nl
	 &         	     135 &  17.9 &  16.6                 & 1.08 &     & ~~~3.3  \nl
K 3-58	 &         	    ~~~8 &  ~8.3 &  ~7.0                 & 1.18 & 6.1 & ~~~5.9  \nl
NGC 6881 & ~~45\tablenotemark{a} &  ~9.6 &  ~4.5                 & 2.14 & 3.1 & ~~11.8  \nl
         & ~~45\tablenotemark{b} &  11.5 & ~4.0\tablenotemark{c} & 2.90 &     & ~~17.4  \nl
M 4-17	 &        	    ~~27 &  20.5 &  16.6                 & 1.23 & 2.2 & ~~~6.4  \nl
K 4-55	 &        	    ~~~4 &  ~7.7 &  ~7.0                 & 1.09 & 1.4 & ~~~0.7  \nl
         &        	    ~~80 &  10.9 &  ~9.0                 & 1.21 &     & ~~~1.8  \nl
M 2-52	 &        	    ~~62 &  ~7.0 &  ~7.0                 & 1.00 & 3.3 &  $<0.3$ \nl
         &        	     120 &  11.5 &  10.9                 & 1.06 &     & ~~~1.5  \nl
\enddata			         
\tablenotetext{a}{Offset of 10 arcsec to the south-east.}
\tablenotetext{b}{Offset of 10 arcsec to the north-west.}
\tablenotetext{c}{FWHM of the line profile.}
\end{deluxetable}

\clearpage

\begin{figure}
\includegraphics{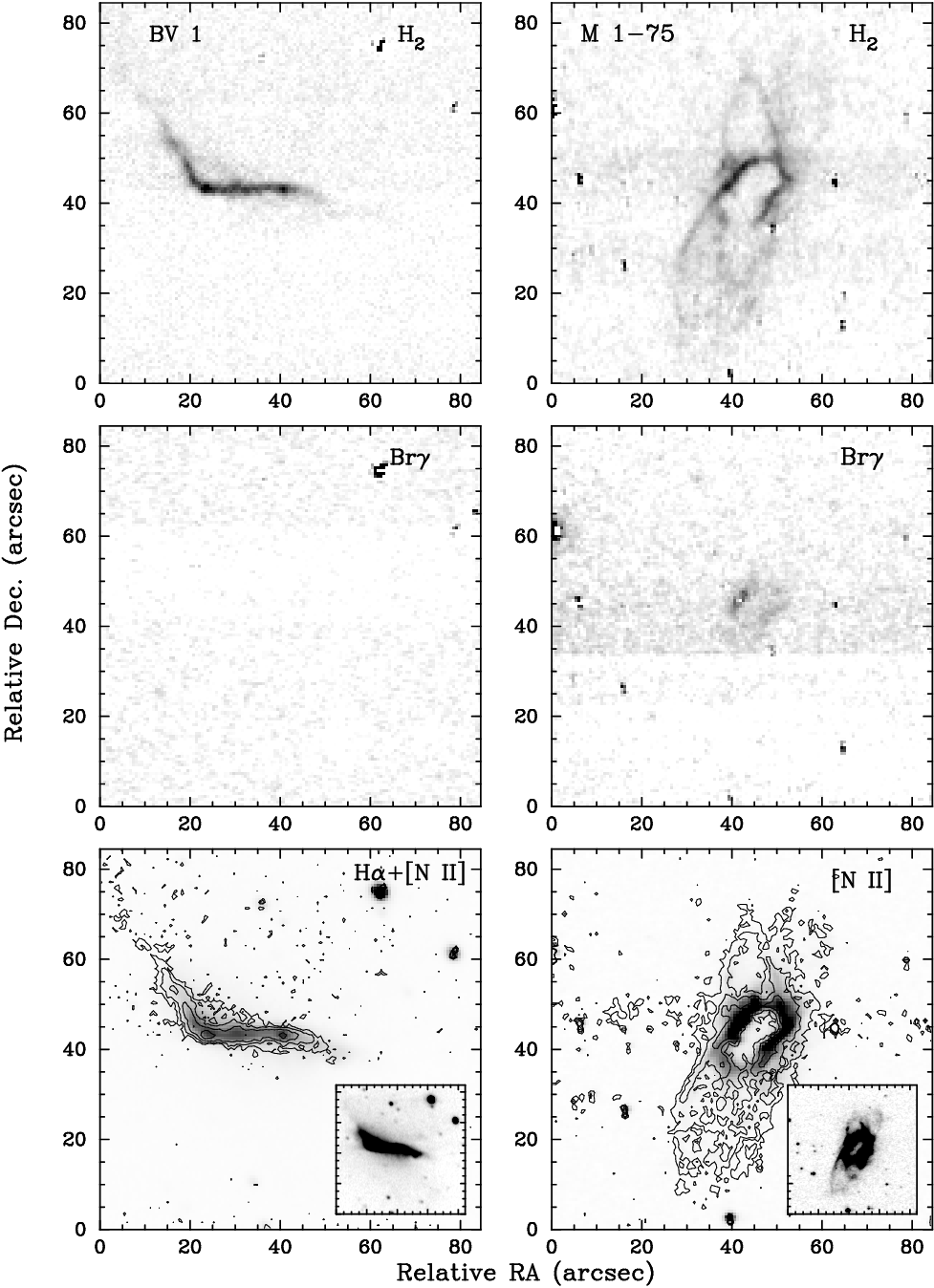}
\end{figure}
     
{\small
\vspace*{20.75cm}
{\sc Fig.~1}{\it a}-- 
Gray-scale representation of the H$_2$ {\it (top)}, and Br$\gamma$ 
{\it (center)} continuum-subtracted images, and of the [N~{\sc ii}] 
emission line image {\it (bottom)} from the Manchado et al.'s catalog 
of BV\,1 {\it (left)}, and M\,1-75 {\it (rigth)}. 
Contour plot of the 2.122 $\mu$m {$\rm H_2$} emission is overlaid on 
the 
[N~{\sc ii}] image whose intensity levels have been set up to allow 
a fair comparison with the molecular emission. 
The lowest contour is at 3$\sigma$. 
The insets at the corners of the [N~{\sc ii}] images are displayed at 
low intensity levels to emphasize the fainter regions of the nebula. 
The image scale is the same for all the images of each nebula. 
North is top, east is left. }

\clearpage

\begin{figure}
\includegraphics{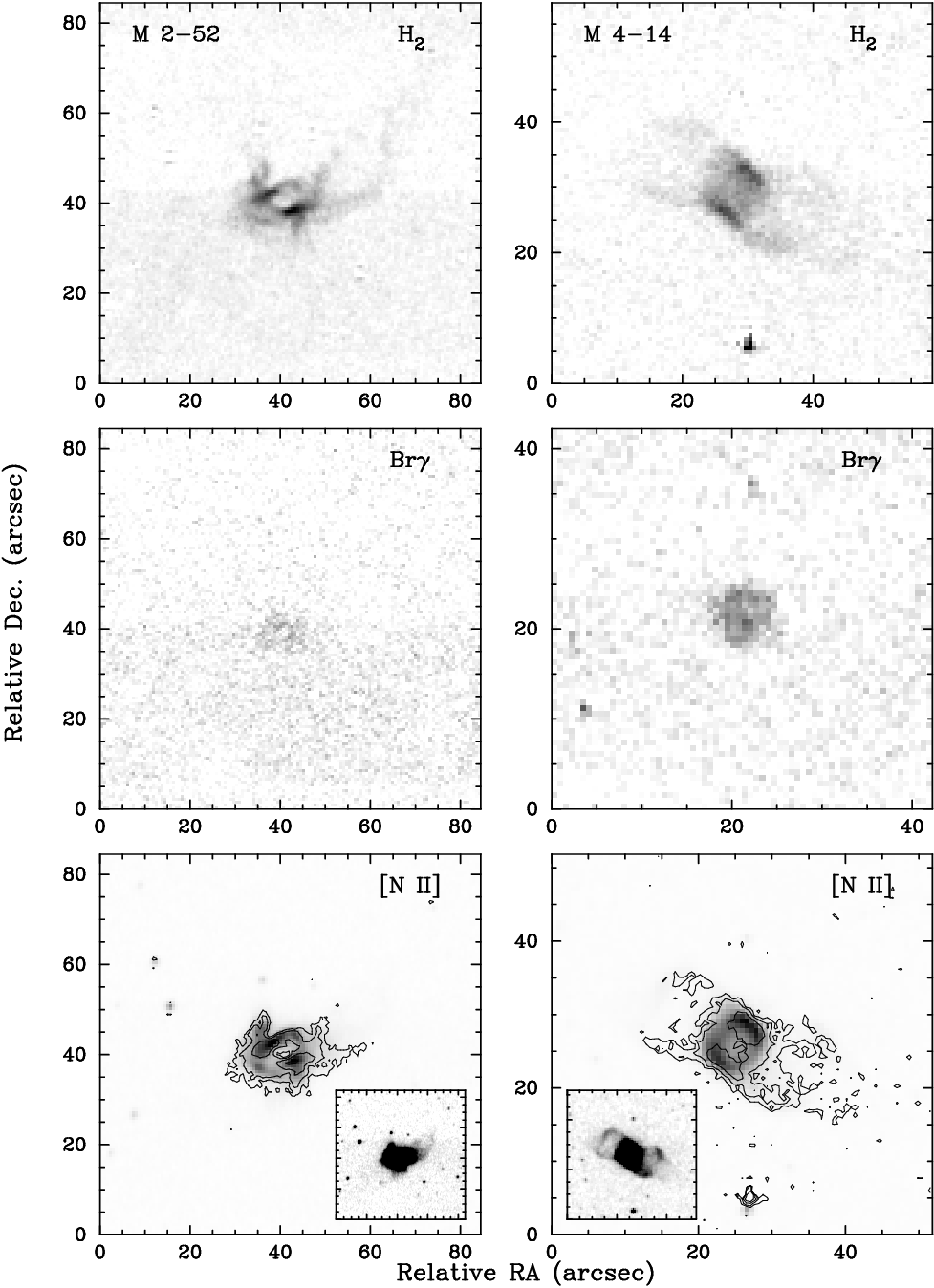}
\end{figure}
     
{\small
\vspace*{20.75cm}
\centerline {{\sc Fig.~1}{\it b}-- 
Same as Fig.~1{\it a}, but for M\,2-52 {\it (left)}, and M\,4-14 {\it (right)}.}}

\clearpage

\begin{figure}
\includegraphics{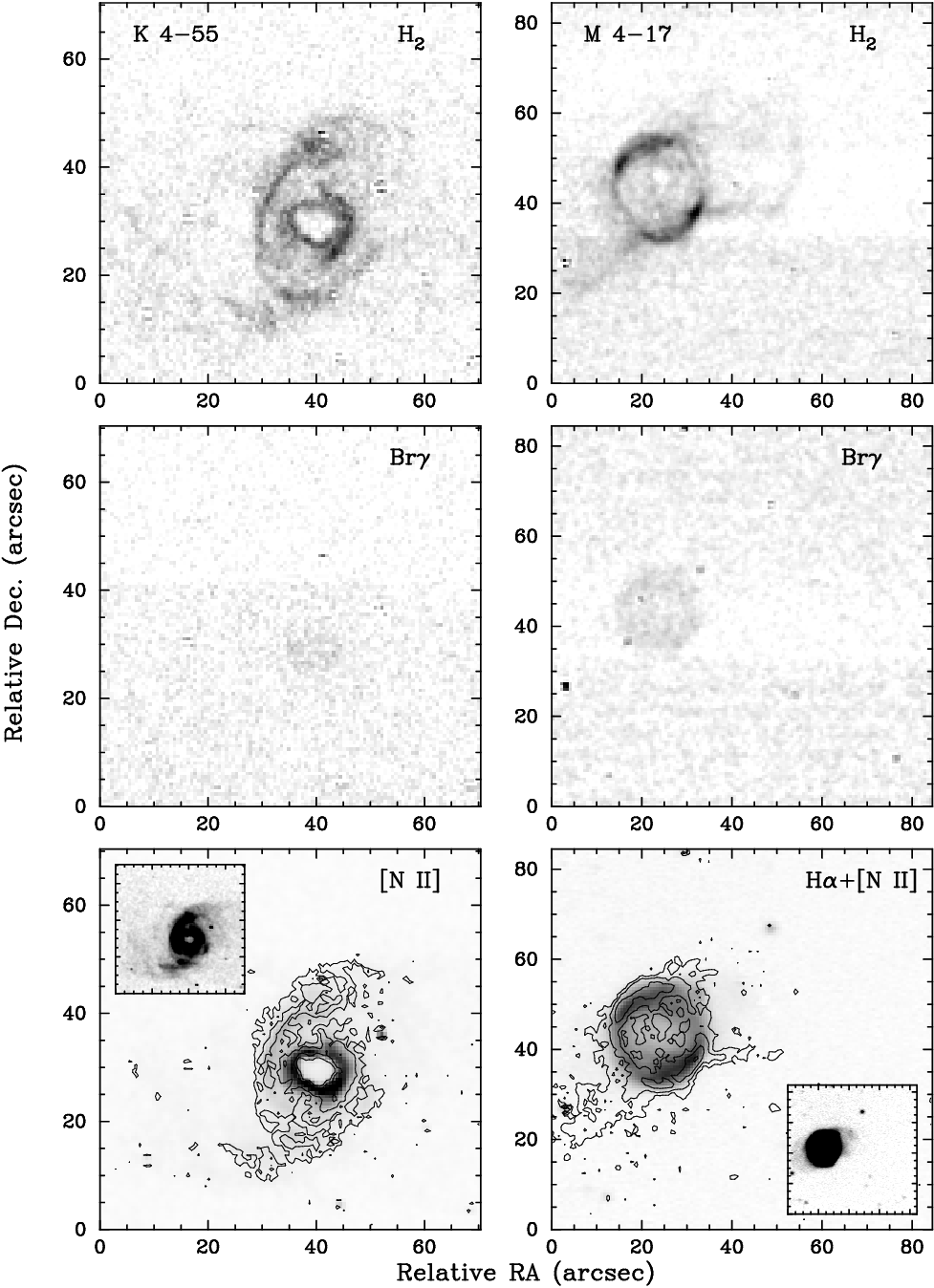}
\end{figure}
     
{\small
\vspace*{20.75cm}
\centerline {{\sc Fig.~1}{\it c}-- 
Same as Fig.~1{\it a}, but for K\,4-55 {\it (left)}, and M\,4-17 {\it (right)}.}}

\clearpage

\begin{figure}
\includegraphics{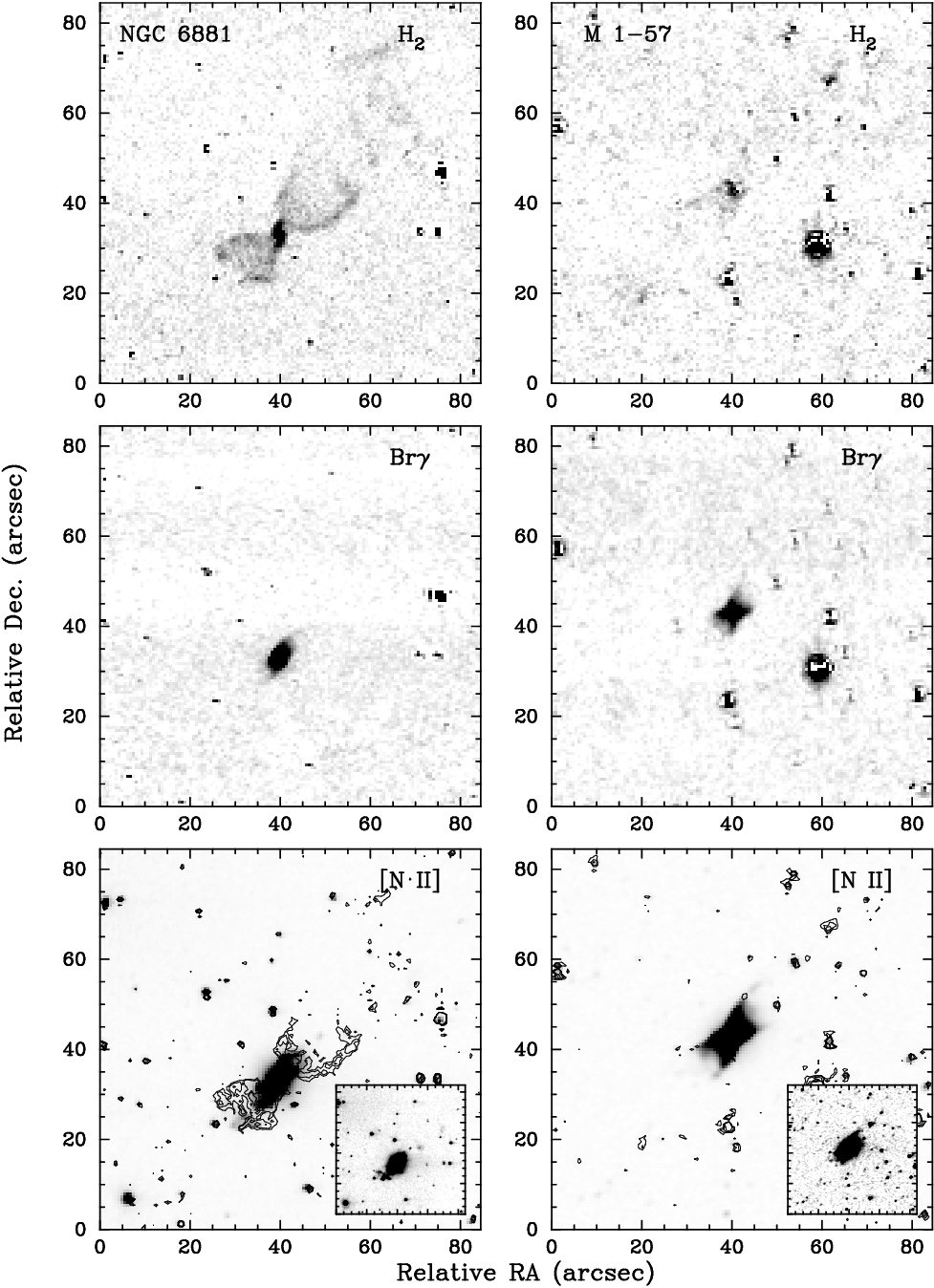}
\end{figure}
     
{\small
\vspace*{20.75cm}
\centerline {{\sc Fig.~1}{\it d}-- 
Same as Fig.~1{\it a}, but for NGC\,6881 {\it (left)}, and M\,1-57 {\it (right)}.}}

\clearpage

\begin{figure}
\includegraphics{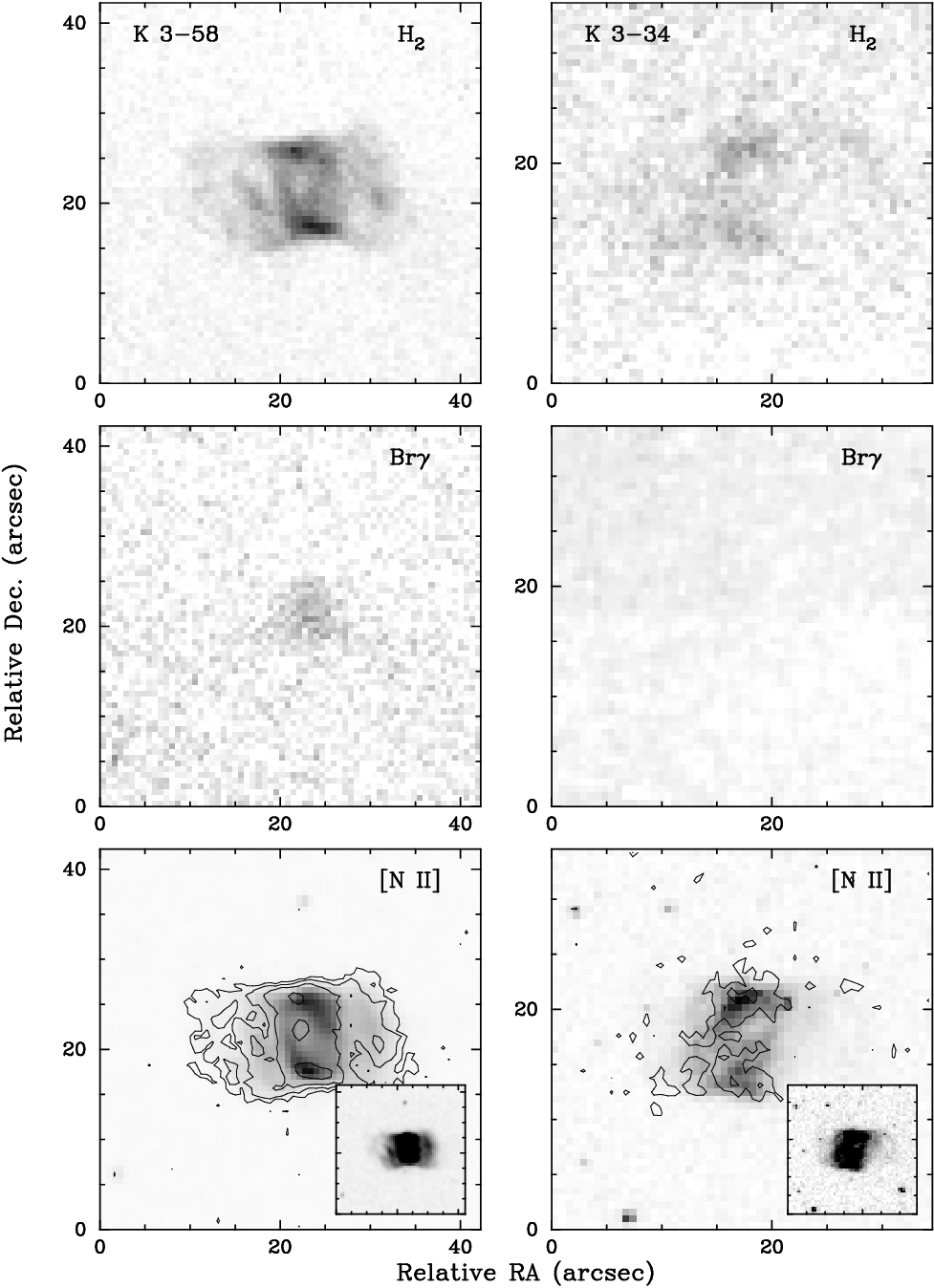}
\end{figure}
     
{\small
\vspace*{20.75cm}
\centerline {{\sc Fig.~1}{\it e}-- 
Same as Fig.~1{\it a}, but for K\,3-58 {\it (left)}, and K\,3-34 {\it (right)}.}}

\clearpage

\begin{figure}
\includegraphics{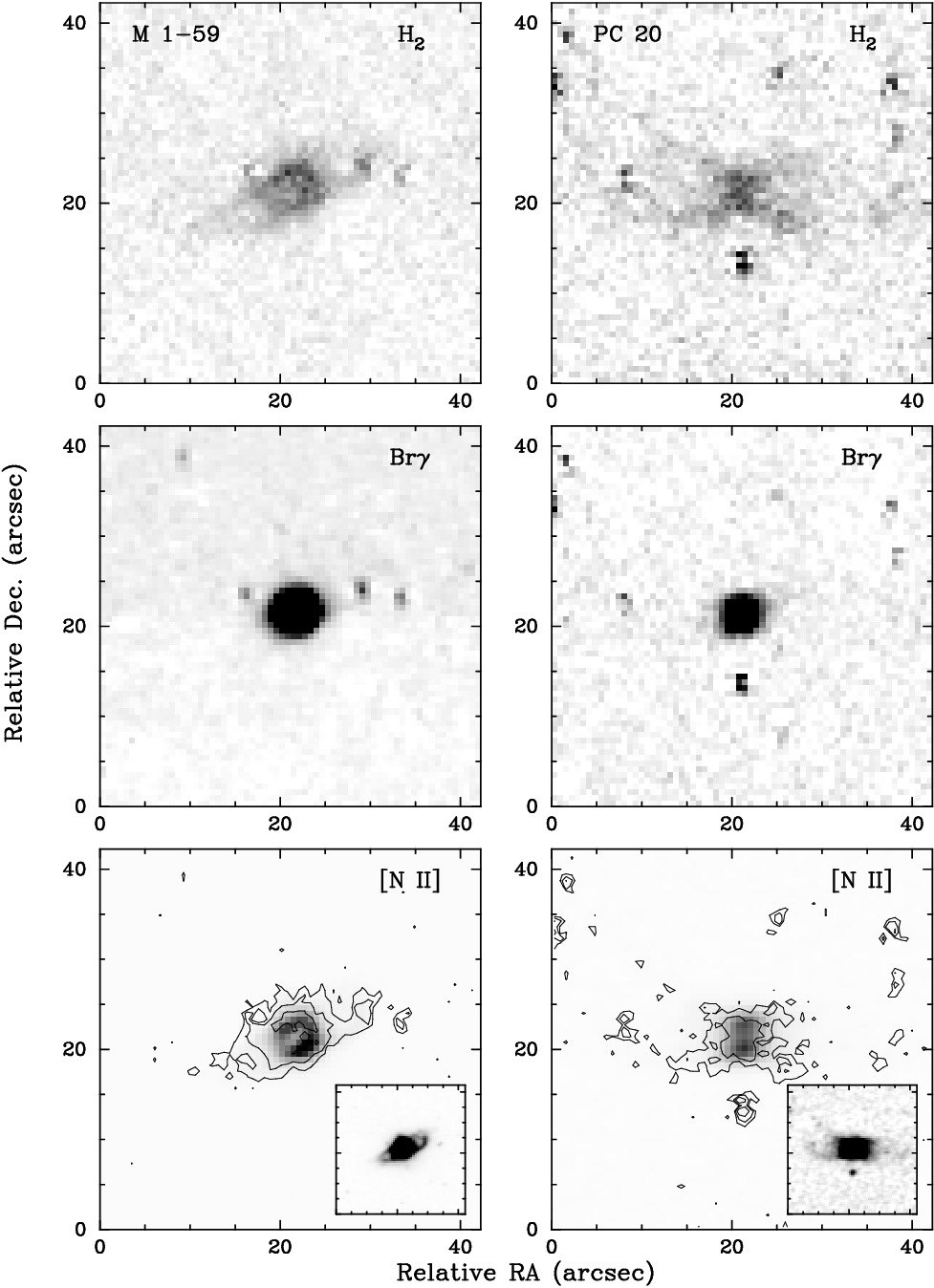}
\end{figure}
     
{\small
\vspace*{20.75cm}
\centerline {{\sc Fig.~1}{\it f}-- 
Same as Fig.~1{\it a}, but for M\,1-59 {\it (left)}, and PC\,20 {\it (right)}.}}

\clearpage

\begin{figure}
\includegraphics{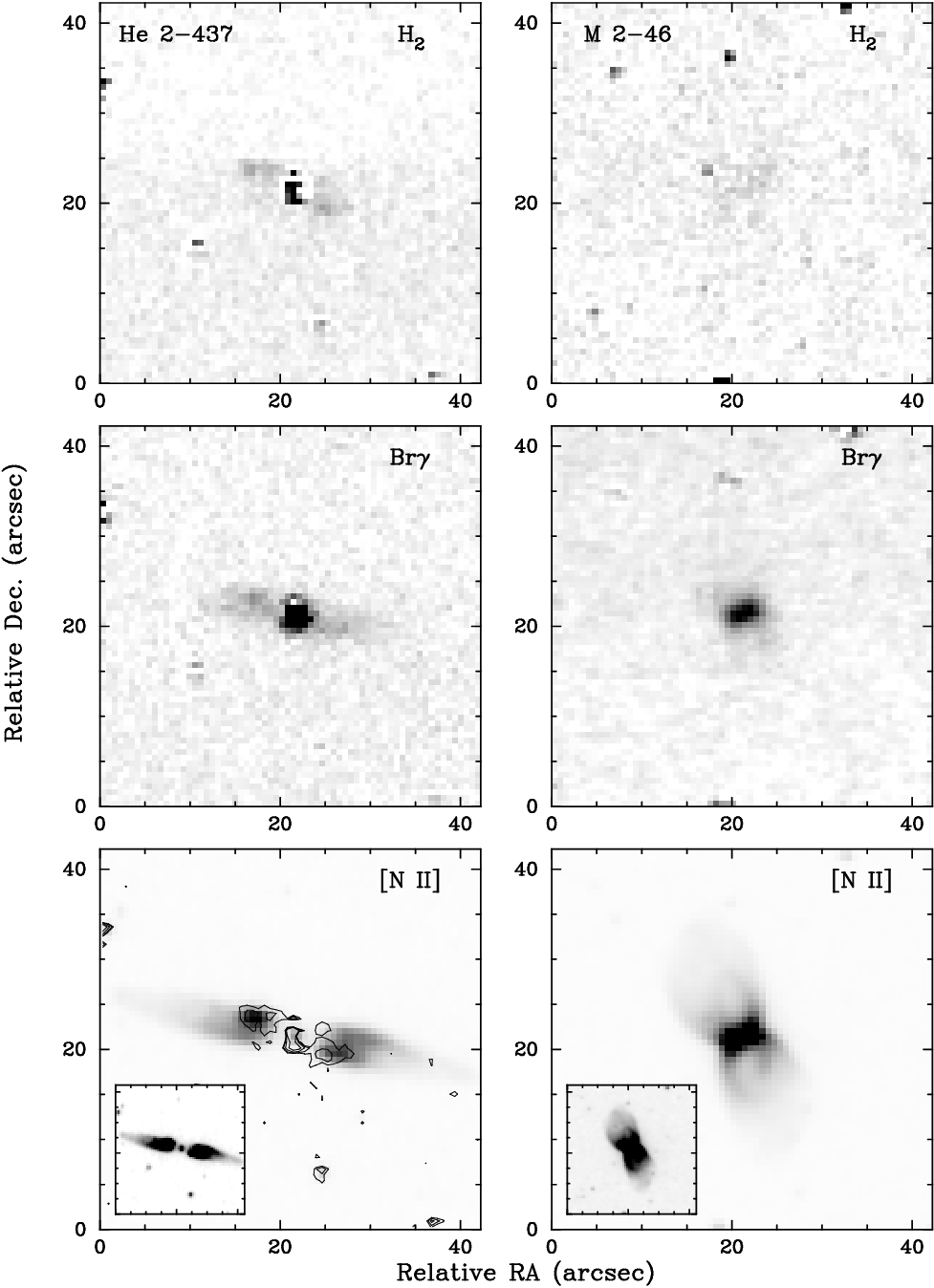}
\end{figure}
     
{\small
\vspace*{20.75cm}
\centerline {{\sc Fig.~1}{\it g}-- 
Same as Fig.~1{\it a}, but for He\,2-437 {\it (left)}, and M\,2-46 {\it (right)}.}}

\clearpage

\begin{figure}
\includegraphics{f1h.eps}
\end{figure}
     
{\small
\vspace*{20.75cm}
\centerline {{\sc Fig.~1}{\it h}-- 
Same as Fig.~1{\it a}, but for He~2-428.}}

\clearpage

\begin{figure}
\includegraphics{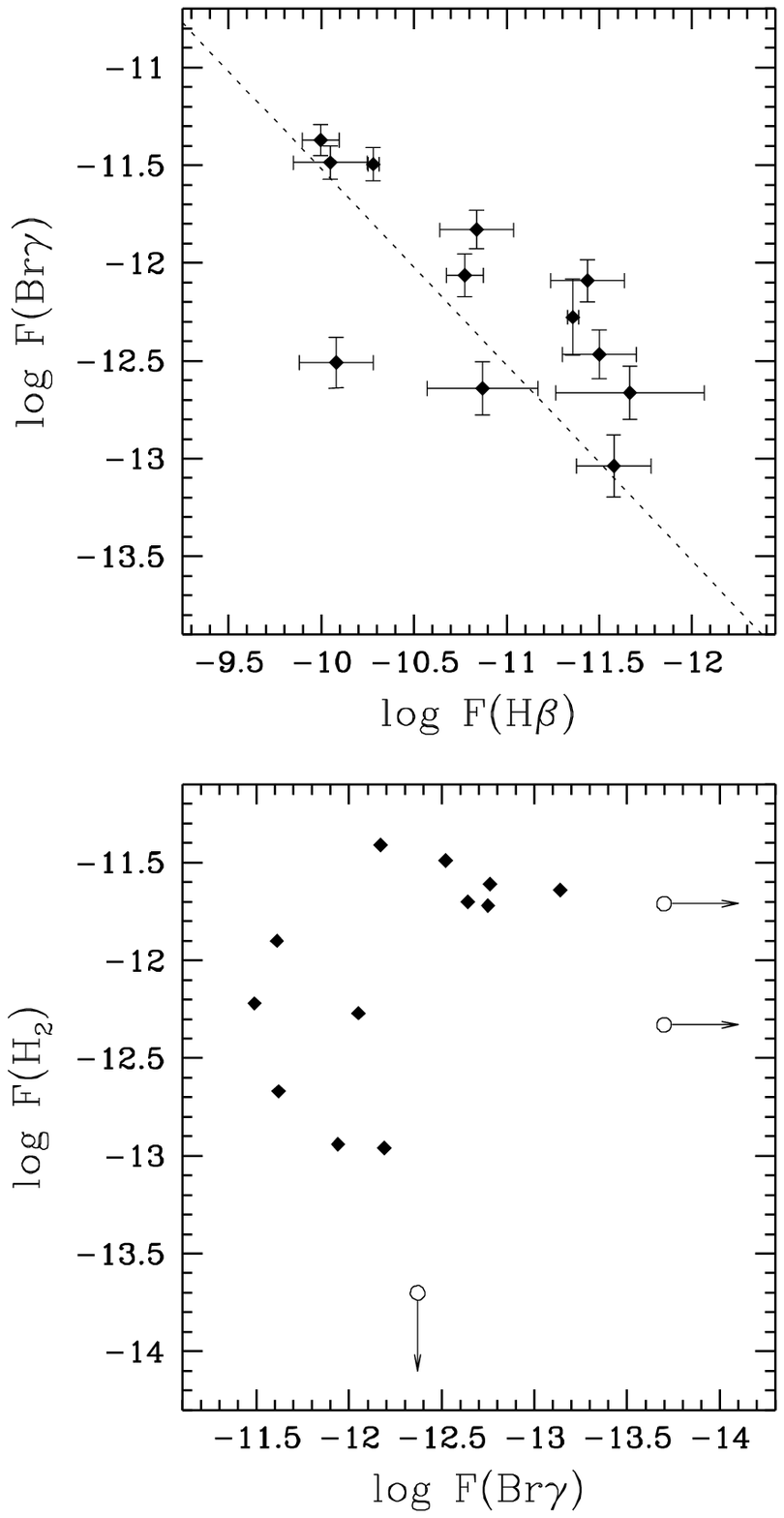}
\end{figure}
     
{\small
\vspace*{17.0cm}
{\sc Fig.~2--} 
{\it (top)} The Br$\gamma$ fluxes plotted against the H$\beta$ fluxes 
as given by Acker at al. (1992). 
The fluxes have been corrected of interstellar extinction using the 
standard extinction law from Pottasch (1984), and the H$\beta$ 
extinction coefficients computed from the comparison of the H$\alpha$ 
to H$\beta$ flux ratio given in Acker (1992) and the theoretical 
Balmer decrement (Osterbrock 1989). 
The errorbar associated to the Br$\gamma$ and H$\beta$ fluxes 
have been plotted, as well as the predicted theoretical ratio 
(dashed line). 
{\it (bottom)} The measured H$_2$ fluxes plotted against the Br$\gamma$ 
fluxes.
When no detection was made, an upper limit of the flux is given and 
indicated by an open dot. 
The sample appears separated in two regions of this diagram 
characterized by stronger H$_2$ or Br$\gamma$ fluxes. }
 
\clearpage
 
\begin{figure}
\includegraphics{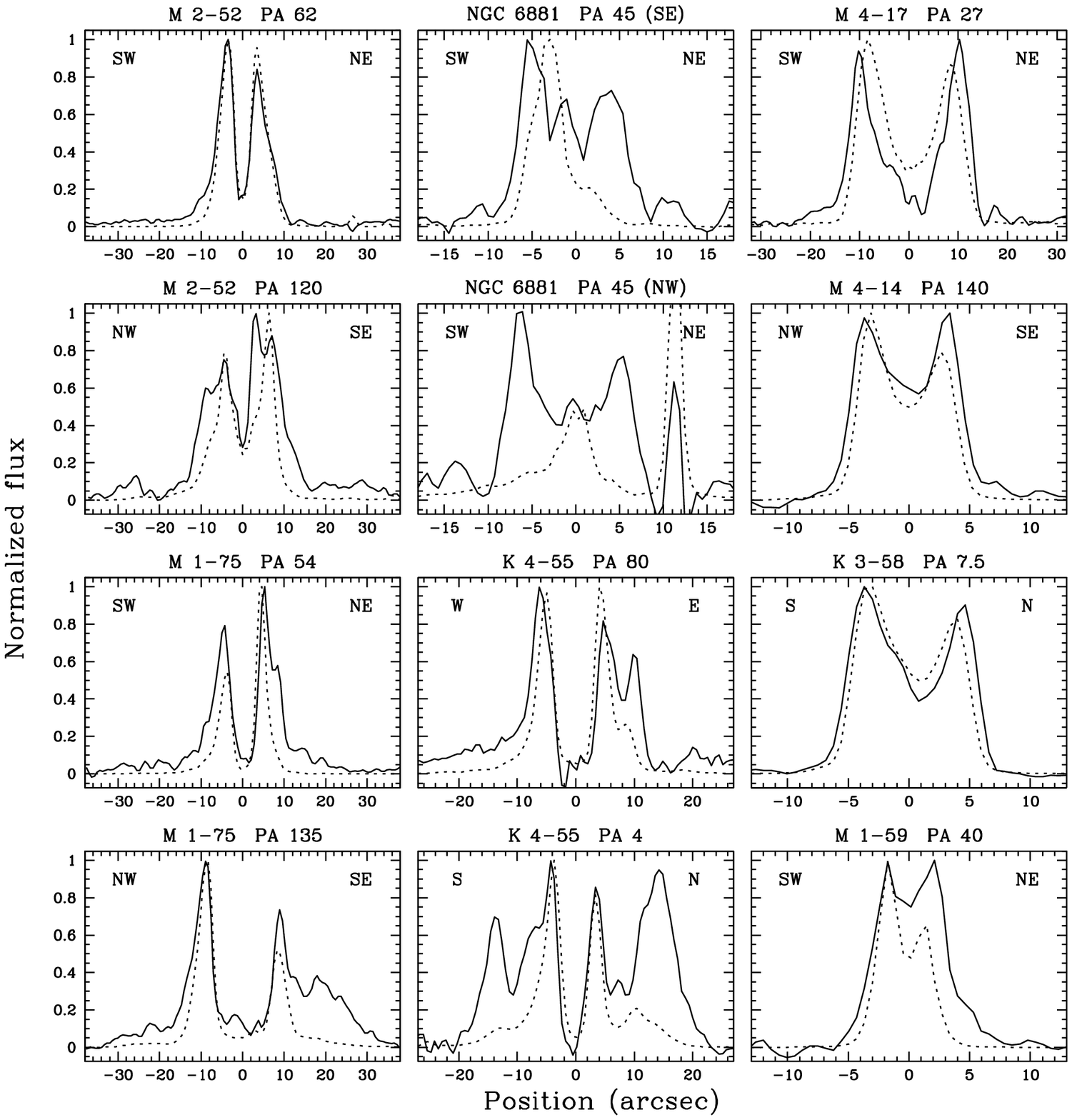}
\end{figure}
     
{\small
\vspace*{18.0cm}
{\sc Fig.~3--} 
Spatial profiles extracted along selected directions comparing the 
2.122 $\mu$m H$_2$ (solid line) and [N~{\sc ii}] optical (dashed 
line) emission. 
The position angle and orientations are indicated on each panel. }

\end{document}